\documentstyle[epsfig]{article}

\title{Searching for WIMP Dark Matter: The case for Germanium Ionization Detectors\footnote{Invited talk given at the 29th
International Meeting on Fundamental Physics, Sitges (Barcelona),
Feb.2001, to be published in the Proceedings.}}

\author{Angel Morales \\ {Laboratory of Nuclear and High Energy Physics} \\
        {University of Zaragoza}  \\
        {50009 Zaragoza. Spain}}
\date{}

\begin{document}


\maketitle

\begin{abstract}
An overview of the main strategies followed in the search for
non-baryonic particle dark matter in the form of WIMPs is given.
To illustrate these searches the case for Germanium ionization
detectors is selected.
\end{abstract}
\section{Introduction}

Experimental observations and robust theoretical arguments have
established that our universe is essentially non-visible, the
luminous matter scarcely accounting for one per cent of the
critical density of a flat universe ($\Omega=1$). The current
picture describes an universe consisting of unknown species of
Dark Energy ($\Omega_{\Lambda}\sim 70\%$) and Dark Matter
($\Omega_{M}\sim 25-30\%$) of which less than $\sim 5\%$ is of
baryonic origin. Most of the Dark Matter would then be made of
non-baryonic particles filling the galactic halos, at least
partially, according to a variety of models. Weak Interacting
Massive (and neutral) Particles (WIMPs) are favourite candidates
to such non-baryonic components. The lightest stable particles of
supersymmetric theories, like the neutralino, describe a
particular class of WIMPs. Without entering into considerations
about how about how large the baryonic dark component of the
galactic halo could be, we take for granted that there is enough
room for WIMPs in our halo, to try to detect them, either directly
or through their by-products. Discovering this form of Dark Matter
is one of the big challenges in Cosmology, Astrophysics and
Particle Physics.

WIMPs can be looked for either directly or indirectly. The
indirect detection of WIMPs proceeds currently through two main
experimental lines: either by looking in cosmic rays experiments
for positrons, antiprotons, or other antinuclei produced by the
WIMPs annihilation in the halo (CAPRICE, BESS, AMS, GLAST,
VERITAS, MAGIC...), or by searching in large underground detectors
(SUPERKAMIOKANDE, SNO, SOUDAN, MACRO) or underwater neutrino
telescopes (BAIKAL, AMANDA, ANTARES, NESTOR) for upward-going
muons produced by the energetic neutrinos emerging as final
products of the WIMPs annihilation in celestial bodies (Sun,
Earth...). This talk will deal with the direct detection of WIMPs.

The direct detection of WIMPs relies in the measurement of the
WIMP elastic scattering off the target nuclei of a  suitable
detector. Pervading the galactic halos, slow moving ($\sim300$
km/s), and heavy ($10 \sim 10^3$ GeV) WIMPs could make a nucleus
recoil with a small energy of a few keV ($\rm T \sim (1-100)$
keV), at a rate which depends of the type of WIMP and interaction.
Only a fraction of the recoil, QT, is visible in the detector,
depending on the type of detector and target and on the mechanism
of energy deposition. The so-called Quenching Factor, Q, is
essentially unit in thermal detectors whereas for the nuclei used
in conventional detectors it ranges from about 0.1 to 0.6. For
instance for a Ge nucleus only about 1/4 of the recoil energy goes
to ionization. On the other hand, the smallness of the
neutralino-matter interaction cross-section makes the rates of the
nuclear recoil looked for very small. The variety of models and
parameters used to describe the Astrophysics, Particle Physics and
Nuclear Physics aspects of the process make the neutralino-nucleus
interaction rate encompass several orders of magnitude, going from
10 to $10^{-5}$ c/kg day, according to the SUSY model and
parameters\cite{treinta}. In fact, it is not higher than $10^{-2}$
c/kg day, for the neutralino parameters which provide the most
favourable relic density ($\rm \Omega\ h^{2}$). Due to such small
rate and small energy deposition produced by the WIMP elastic
scattering, the direct search for particle dark matter through
their scattering by nuclear targets requires ultralow background
detectors of a very low energy threshold. Moreover, the (almost)
exponentially decreasing shape of the predicted nuclear recoil
spectrum mimics that of the low energy background registered by
the detector. All these features together make the WIMP detection
a formidable experimental challenge.

Customarily, one compares the predicted event rate with the
observed spectrum. If the former turns out to be larger than the
measured one, the particle which would produce such event rate can
be ruled out as a Dark Matter candidate. That is expressed as a
contour line $\sigma$(m) in the plane of the WIMP-nucleon elastic
scattering cross section versus the WIMP mass. That excludes, for
each mass m, those particles with a cross-section above the
contour line $\sigma$(m). The level of background sets,
consequently, the sensitivity of the experiment in eliminating
candidates or in constraining their masses and cross sections.

However, this mere comparison of the expected signal with the
experimentally observed spectrum will not be able to detect the
WIMP, because such spectrum, even extremely small, could still be
due to pure background sources. In other words, it would be
difficult to be convinced that such signal is due to WIMPs and not
to mere background sources. A convincing proof of the detection of
WIMPs would need to find unique signatures in the data
characteristic of them and not of the background. There exist some
temporal or spatial asymmetries specific of the WIMP interaction,
which cannot be faked by the background or by instrumental
artifacts. They are due to the kinematics of the motion of the
Earth (and of our detectors) in the galactic halo. The only
distinctive signature investigated up to now is a predicted annual
modulation of the WIMP signal rate due to the seasonal variation
produced by the Earth's motion with respect to the Sun. Such a
seasonal affect has been found by the DAMA experiment at the
$3\sigma$ level and has been associated to the existence of a
WIMP.

The detectors used so far are: ionization detectors of Ge (IGEX,
COSME, H/M, HDMS) and of Si (UCSB), scintillation crystals of NaI
(ZARAGOZA, DAMA, UKDMC, SACLAY, ELEGANTS), liquid or liquid-gas
Xenon detectors (DAMA, UCLA, UKDMC), calcium fluoride
scintillators (MILAN, OSAKA, ROMA), thermal detectors (bolometers)
with saphire absorbers (CRESST, ROSEBUD), with telurite absorbers,
(MIBETA, CUORICINO) or with germanium absorbers (ROSEBUD) as well
as bolometers which also measure the ionization, like that of Si
(CDMS) and of Ge (CDMS, EDELWEISS). New detectors and techniques
are entering the stage. Worth to be mentioned are: scintillating
bolometers of calcium tungstate which measure heat and high
(CRESST and ROSEBUD), and of BGO (ROSEBUD); a TPC sensitive to the
direction of the nuclear recoil (DRIFT); devices which use
superheated droplets (SIMPLE and PICASSO), or those which use
colloids of superconducting superheated grains (ORPHEUS). There
exist also projects featuring a large amount of target nuclei in
segmented detectors, both with ionization Ge detectors (GENIUS,
GEDEON) and cryogenic thermal devices (CUORE). Table 4 gives an
overview of the experiments on direct detection of WIMPs currently
in operation or in preparation. A review of neutralino dark matter
can be found in Ref \cite{Gri}. WIMP direct detection are
reviewed, for instance in Ref \cite{Mor} and \cite{Mor2}. Recent
results are given in Ref \cite{Mor3}.

\section{Strategies for WIMP detection}

The smallness and rarity of WIMP signals dictate the experimental
strategies for their detection: reduce first the background, by
controlling the radiopurity of the detector, its components, the
shielding and the environment. The best radiopurity has been
obtained in conventional Ge experiments (IGEX, H/M). In the case
of the NaI scintillators (DAMA, UKDM, ANAIS), the achieved
backgrounds are still one order of magnitude worse than those
reached in Ge, but the use of new radiopure powder in the crystals
has lead to substantial improvements (DAMA). The next step is to
use discrimination mechanisms able to distinguish electron recoils
(tracers of the background) from nuclear recoils (originated by
WIMPs or neutrons). Two types of techniques have been applied for
such purpose: a statistical pulse shape analysis (PSD) based on
the different timing behaviour of both types of pulses and a
background rejection method based on the identification (on an
event-by-event basis) of the nuclear recoils, by measuring at the
same time two different mechanisms of energy deposition, like the
ionization (or scintillation) and heat, capitalizing the fact that
for a given deposited energy (measured as phonons) the recoiling
nucleus ionizes less than the electrons. Examples of PSD are
UKDMC, Saclay, DAMA and ANAIS, whereas the event by event
discrimination has been successfully applied in CDMS and EDELWEISS
by measuring ionization and heat and now in CRESST and ROSEBUD by
measuring light and heat.

A promising discriminating technique is that used in the two-phase
liquid-gas Xenon detector with ionization plus scintillation, of
the ZEPLIN series of detectors. An electric field prevents
recombination, the charge being drifted to create a second pulse
in addition to the primary pulse. The amplitudes of both pulses
are different for nuclear recoils and gammas allowing their
discrimination.

Another technique is to discriminate gamma background from
neutrons (and so WIMPs) using threshold detectors---like neutron
dosimeters---which are blind to most of the low Linear Energy
Transfer (LET) radiation (e, $\mu$, $\gamma$). Detectors which use
superheated droplets which vaporize into bubbles by the WIMP (or
other high LET particles) energy deposition are SIMPLE and
PICASSO. An ultimate discrimination will be the identification of
the different kind of particles by the tracking they left in, say,
a TPC, plus the identification of the WIMP through the directional
sensitivity of the device (DRIFT).

Obviously one should try to make detectors of very low energy
threshold and high efficiency to see most of the signal spectrum,
not just the tail. That is the case for the bolometer experiments
which seen efficiently the energy delivered by the WIMP (quenching
factor is unity), (MIBETA, CRESST, ROSEBUD, CUORICINO, CDMS and
EDELWEISS).

After the process of reducing and identifying the background is
driven to its best, then one should search for distinctive
signatures of the dark matter particles to prove that you are,
indeed, seeing a WIMP. Identifying labels are derived from the
Earth motion through the galactic halo, which produces two
assymetries distinctive of WIMPs: First, the Earth orbital motion
around the sun has a summer-winter variation which results in a
small (5\%) annual modulation of the WIMP interaction rates; and
second, the Earth (and solar system) motion through the galactic
centre produces a large ($\sim 1$) directional assymetry
forward-backward of the recoiling nucleus. The annual modulation
signature has been already explored. Pioneering searches for WIMP
annual modulation signals were carried out in Canfranc (NaI-32),
Kamioka (ELEGANTS) and Gran Sasso (DAMA-Xe). Recently the DAMA
experiment at Gran Sasso, using a set of NaI scintillators
reported in 1997 and 2000 an annual modulation effect of at the
$3\sigma$ level interpreted as due to a WIMP of about 60 GeV of
mass and scalar cross-section on protons of $\sigma_{\rm p} = 7
\times 10^{-6}$ picobarns. As far as the forward-backward
assymetry of the signal is concerned the DRIFT project will search
for it. Information and current results of the experiments
mentioned above can be found in References \cite{Mor3} and
\cite{taup}.

In particular, in the underground  facility of the Canfranc Tunnel
(Spain), various of  the techniques mentioned above are currently
employed in a search for WIMPs\cite{Ceb00}: Ge ionization
detectors, (COSME and IGEX), Sodium iodine scintillators (NaI-32
and ANAIS), thermal detectors with Al$_{2}$O$_{3}$ and Ge
absorbers (ROSEBUD-I) as well as and with calcium tungstate and
BGO scintillating bolometers (ROSEBUD-II). To illustrate how these
searches proceed we will present the case for Ge detectors, and
will describe its status, achievements and results. As one of the
main achievements in Ge experiments is their low radioactive
background, essential ingredient in the search for dark matter, we
will illustrate it with a saga of Germanium experiments performed
in Canfranc and the succesive suppresion of the background
obtained. Such reduction has provided the lowest raw background
rate obtained so far and, consequently, the stringest exclussion
plots ever derived with ionization detectors.

\begin{figure}[ht]
\centerline{ \epsfxsize=10cm \epsffile{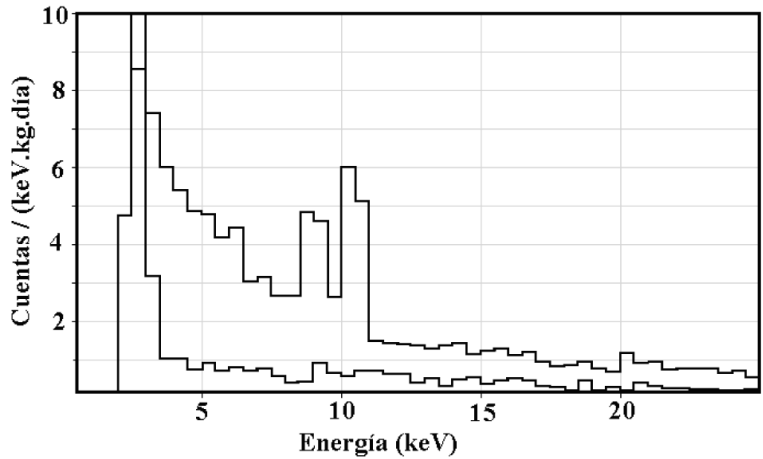}}
 \caption{Low energy background spectra of COSME-1 and COSME-2.}
 \label{fig_1}
\end{figure}

\section{Germanium Experiments}

The high radiopurity and low background achieved in Germanium
detectors, their fair low energy threshold, their reasonable
Quenching Factor (about 25\%) (nuclear recoil ionization
efficiency relative to that of electrons of the same kinetic
energy, or ionization yield) and other nuclear merits make
Germanium a good option to search for WIMPs with detectors and
techniques fully mastered. The first detectors applied to WIMP
direct searches (as early as in 1987) were, in fact, Ge diodes, as
by-products of $2 \beta$-decay dedicated experiments. The
exclusion plots $\sigma$(m) for spin-independent couplings
obtained by former Ge experiments [PNNL/USC/Zaragoza (TWIN
\cite{siete} and COSME-1\cite{Jmor}\cite{Gar92}), UCSB
\cite{diez}, CALT/NEU/PSI \cite{once}, H/M \cite{doce}] are still
remarkable and have not been surpassed till recently by NaI
experiments (DAMA)\cite{trece} using statistical pulse shape
discrimination (PSD) or by cryogenic experiments measuring both
heat and ionization like CDMS\cite{Abusaidi:2000} and
EDELWEISS\cite{Benoit:2001}. Table 5 shows the Germanium
ionization detector experiments currently in operation.

A Ge detector of natural isotopic abundance (COSME) of the U.
Zaragoza / U. S. Carolina / PNNL Collaboration and another one
(RG-II) made of enriched $^{76}$Ge of the IGEX Collaboration are
being used in Canfranc to search for WIMPs interacting coherently
with the Ge nuclei of the detectors.

The COSME detector was fabricated at Princeton Gamma-Tech, Inc. in
Princeton, New Jersey, using germanium of natural isotopic
abundance. The refinement of newly-mined germanium ore to finished
metal for this detector was expedited to minimize production of
cosmogenic $^{68}$Ge. The detector is a p-type coaxial hyperpure
natural germanium crystal with a mass of 254 g and an active mass
of 234 g which has a long term resolution of 0.43 keV full width
at half maximum (FWHM) at the 10.37 keV gallium X-ray. In its
first Canfranc installation (at 675 m.w.e) the detector was placed
within a shielding of 10 cm of 2000 yr. old (Roman) lead (inner
layer) plus 20 cm of low activity lead (about 70 yr old). A 3 mm
thick PVC box sealed with silicone closed the lead shielding to
purge the radon gas. The PVC box was covered by 1 mm of cadmium
and 20 cm of paraffin and borated polyethylene. All the shielding
and mounting was supported by 10 cm of vibrational and acoustic
insulator sandwiched within two layers of 10 cm of wood mounted on
a floor of concrete (20 cm). It was operated in the former
Canfranc underground facility at 675 m.w.e. in a small gallery of
the Canfranc railway tunnel in the Spanish Pyrenees closed to the
traffic. In that set-up, the energy threshold was $\rm E_{thr} =
1.6$ keV and the background at threshold was about 10 counts
keV$^{-1}$ kg$^{-1}$ day$^{-1}$. After about 500 days of data
taking (referred as COSME-1 experiment) no signal originated by
WIMP appeared and, from the background spectrum,  an exclusion
plot in the cross-section versus WIMP masses plane was derived,
assuming WIMP-matter spin-independent couplings. In the COSME-1
data, the energy range chosen to derive the exclusion plot was
from 1.6 to 8 keV where the background was, approximately, 5
counts/(keV kg day). In spite of such modest figure, the results
after 130 kg day of exposure improved the exclusion plots at low
masses (from 9 to 20 GeV) obtained with other Ge experiments
because of the low threshold energy of COSME-1.

The COSME detector has been reinstalled, in better background
conditions in the new Canfranc Underground Laboratory LSC (at 2450
m.w.e.) inside a Marinelli beaker in Roman lead (COSME-2) in a
common shielding (described later) together with the three 2.1 kg
enriched germanium detectors of IGEX (the International Germanium
Experiment on Double Beta Decay). In its new installation, COSME-2
has an energy threshold of $E_{thr}=2.5$ keV, and an energy
resolution of $\Gamma (\rm FWHM) = 0.4$ keV at 10 keV. The average
background rate, in 311 days of exposure (Mt=72.8 kg day) is 0.6
c/(keV kg day) from 2 to 15 keV and 0.3 c/(keV kg day) from 15 to
30 keV which is significantly better (more than one order of
magnitude) than in COSME-1. The COSME-2 spectrum is shown in Fig.
\ref{fig_1} together with that of COSME-1, to show the remarkable
background reduction obtained in the new set-up and shielding and
the disappearance of the cosmogenically induced peaks in the 8-10
keV energy region due to the about ten years elapsed between the
two experiments. The numerical data of COSME 2 are given in Table
\ref{tab-cos-9}.

\begin{table}[h]
\begin{center}
\begin{tabular}[h]{cccccc}
\hline \multicolumn{1}{r}{{\bf E (keV)}}
                 & \multicolumn{1}{r}{{\bf counts}}

                 & \multicolumn{1}{r}{{\bf E (keV)}}
                 & \multicolumn{1}{r}{{\bf counts}}

                 & \multicolumn{1}{r}{{\bf E (keV)}}
                 & \multicolumn{1}{r}{{\bf counts}}
                  \\
\hline
\small
2.5 & 1098 & 18.5 & 25 & 34.5 & 21\\
3.5 & 76 & 19.5 & 19 & 35.5 & 13\\
4.5 & 62 & 20.5 & 28 & 36.5 & 20\\
5.5 & 57 & 21.5 & 20 & 37.5 & 12\\
6.5 & 56 & 22.5 & 18 & 38.5 & 14\\
7.5 & 37 & 23.5 & 17 & 39.5 & 20\\
8.5 & 51 & 24.5 & 22 & 40.5 & 13\\
9.5 & 45 & 25.5 & 20 & 41.5 & 17\\
10.5 & 56 & 26.5 & 23 & 42.5 & 17\\
11.5 & 46 & 27.5 & 28 & 43.5 & 17\\
12.5 & 34 & 28.5 & 19 & 44.5 & 20\\
13.5 & 30 & 29.5 & 19 & 45.5 & 18\\
14.5 & 34 & 30.5 & 19 & 46.5 & 48\\
15.5 & 38 & 31.5 & 15 & 47.5 & 20\\
16.5 & 30 & 32.5 & 27 & 48.5 & 17\\
17.5 & 19 & 33.5 & 22 & 49.5 & 19\\ \hline
\end{tabular}
\caption{Low-energy data from the COSME-2 detector (Mt $=$
73~kg-d).} \label{tab-cos-9}
\end{center}
\end{table}

The International Germanium Experiment (IGEX) is an experiment
which was designed to investigate the double beta decay of
Germanium. It uses three enriched detectors of $^{76}$Ge of
$\sim$2.1 Kg each, and it has been described in detail elsewhere
\cite{Aal}\cite{Gon99}. One of these detectors, RG-II, has been
recently prepared for use at low energies in typical WIMP searches
(referred to as IGEX-DM) .

The IGEX detectors (and so RG-II) were fabricated at Oxford
Instruments, Inc., in Oak Ridge, Tennessee. Russian GeO$_2$
powder, isotopically enriched to 86\% $^{76}$Ge, was purified,
reduced to metal, and zone refined to $\sim 10^{13}$ p-type donor
impurities per cubic centimeter by Eagle Picher, Inc., in Quapaw,
Oklahoma. The metal was then transported to Oxford Instruments by
surface in order to minimize activation by cosmic ray neutrons,
where it was further zone refined, grown into crystals, and
fabricated into detectors. All of the cryostat parts were
electroformed using a high purity OFHC copper/CuSO$_4$/H$_2$SO$_4$
plating system. The solution was continuously filtered to
eliminate copper oxide, which causes porosity in the copper. A
Ba(OH)$_2$ solution  was added to precipitate BaSO$_4$, which is
also collected on the filter. Radium in the bath exchanges with
the barium on the filter, thus minimizing radium contamination in
the cryostat parts. The CuSO$_4$ crystals were purified of thorium
by multiple recrystallization.

The RG-II detector, has a mass of $\sim2.2$ kg. and an active mass
of 2.0 kg, as measured with a collimated source of $^{152}$Eu in
the Canfranc Laboratory, in agreement with the Oxford Instruments
efficiency measurements. The full-width at half-maximum (FWHM)
energy resolution of RG-II is 2.37~keV at the 1333-keV line of
$^{60}$Co. Energy calibration and resolution measurements were
made every 7--10 days using the lines of $^{22}$Na and $^{60}$Co.
Calibration for the low energy region was extrapolated using the
X-ray lines of Pb. The first-stage field-effect transistor (FET)
is mounted on a Teflon block a few centimeters from the central
contact of the germanium crystal. The protective cover of the FET
and the glass shell of the feedback resistor have been removed to
reduce radioactive background. This first-stage assembly is
mounted behind a 2.5-cm-thick cylinder of archaeological lead to
further reduce background. Further stages of preamplification are
located at the back of the cryostat cross arm, approximately 70 cm
from the crystal. The detector has preamplifier modified for the
pulse-shape analysis.

\begin{figure}[t]
\centerline{ \epsfxsize=10cm \epsffile{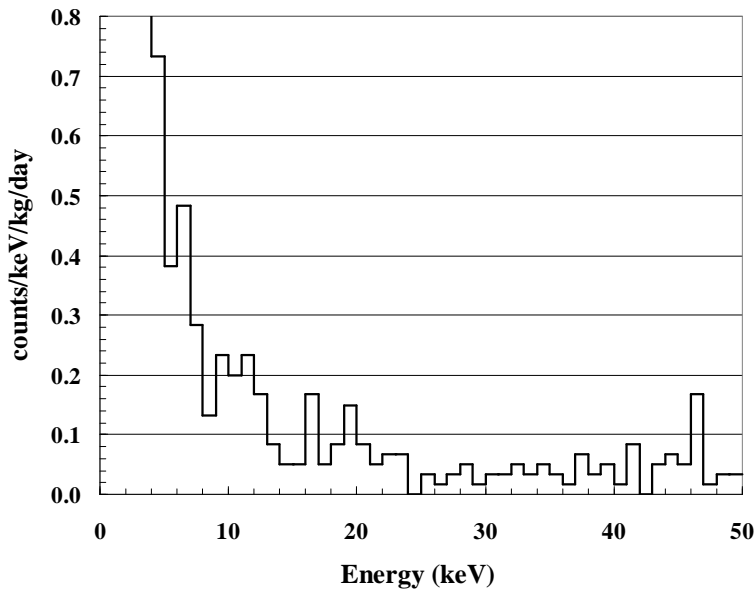} }
 \caption{Low-energy spectrum of the IGEX RG-II detector
(Mt $=$ 60~kg-d), labelled IGEX-2000.}
 \label{dm-ig-2}
\end{figure}

The detectors shielding (shared by COSME, RG-II and two other IGEX
detectors) is as follows, from inside to outside. The innermost
shield consists of 2.5 tons of 2000-year-old archaeological lead
forming a 60-cm cube and having $<9$~mBq/kg of
$^{210}$Pb($^{210}$Bi), $< 0.2$~mBq/kg of $^{238}$U, and
$<0.3$~mBq/kg of $^{232}$Th. The detectors fit into
precision-machined holes in this central core, which minimizes the
empty space around the detectors available to radon. Nitrogen gas,
at a rate of 140~l/hour, evaporating from liquid nitrogen, is
forced into the detector chambers to create a positive pressure
and further minimize radon intrusion. The archaeological lead
block is centered in a 1-m cube of 70-year-old low-activity
lead($\sim 10$ tons) having $\sim 30$~Bq/kg of $^{210}$Pb. A
minimum of 15~cm of archaeological lead separates the detectors
from the outer lead shield. A 2-mm-thick cadmium sheet surrounds
the main lead shield, and two layers of plastic seal this central
assembly against radon intrusion. A cosmic muon veto covers the
top and sides of the central core, except where the detector
Dewars are located. The veto consists of BICRON BC-408 plastic
scintillators 5.08 cm $\times$ 50.8 cm $\times$ 101.6 cm with
surfaces finished by diamond mill to optimize internal reflection.
BC-800 (UVT) light guides on the ends taper to 5.08 cm in diameter
over a length of 50.8 cm and are coupled to Hamamatsu R329
photomultiplier tubes. The anticoincidence veto signal is obtained
from the logical OR of all photomultiplier tube discriminator
outputs. An external polyethylene neutron moderator 20~cm thick
(1.5 tons) completes the shield. The entire shield is supported by
an iron structure resting on noise-isolation blocks. The
experiment\cite{Mor00} (referred to as IGEX-2000) is located in a
room isolated from the rest of the laboratory and has an
overburden of 2450 m.w.e., which reduces the muon flux down to a
measured $2 \times 10^{-7} \rm cm^{-2} \rm s^{-1}$.

\begin{table}[h]
\begin{center}
\begin{tabular}[h]{cccccc}
\hline \multicolumn{1}{r}{{\bf E (keV)}}
                 & \multicolumn{1}{r}{{\bf counts}}

                 & \multicolumn{1}{r}{{\bf E (keV)}}
                 & \multicolumn{1}{r}{{\bf counts}}

                 & \multicolumn{1}{r}{{\bf E (keV)}}
                 & \multicolumn{1}{r}{{\bf counts}}
                  \\
\hline
\small
4.5 & 44 & 19.5 & 9 & 34.5 & 3\\ 5.5 & 23 & 20.5 & 5 & 35.5 & 2\\
6.5 & 29 & 21.5 & 3 & 36.5 & 1\\ 7.5 & 17 & 22.5 & 4 & 37.5 & 4\\
8.5 & 8 & 23.5 & 4 & 38.5 & 2\\ 9.5 & 14 & 24.5 & 0 & 39.5 & 3\\
10.5 & 12 & 25.5 & 2 & 40.5 & 1\\ 11.5 & 14 & 26.5 & 1 & 41.5 &
5\\ 12.5 & 10 & 27.5 & 2 & 42.5 & 0\\ 13.5 & 5 & 28.5 & 3 & 43.5 &
3\\ 14.5 & 3 & 29.5 & 1 & 44.5 & 4\\ 15.5 & 3 & 30.5 & 2 & 45.5 &
3\\ 16.5 & 10 & 31.5 & 2 & 46.5 & 10\\ 17.5 & 3 & 32.5 & 3 & 47.5
& 1\\ 18.5 & 5 & 33.5 & 2 & 48.5 & 2\\ \hline
\end{tabular}
\caption{Low-energy data from the IGEX RG-II detector (Mt $=$
60~kg-d), labelled IGEX-2000.} \label{tab-ig-10}
\end{center}
\end{table}

The data acquisition system is based on standard NIM electronics.
It has been implemented by splitting the normal preamplifier
output pulses of each detector and routing them through two
Canberra 2020 amplifiers having different shaping times enabling
noise rejection as first applied in Ref \cite{Jmor}. These
amplifier outputs are converted using 200 MHz Wilkinson-type
Canberra analog-to-digital converters, controlled by a PC through
parallel interfaces. For each event, the arrival time (with an
accuracy of 100~$\mu$s), the elapsed time since the last veto
event (with an accuracy of 20~$\mu$s), and the energy from each
ADC are recorded. For a total counting rate smaller than 1 Hz, the
dead time is negligible. The muon veto anticoincidence was done
off-line with a software window of 240~$\mu$s.

For the data taken in a new set-up, described below, (and referred
to as IGEX-2001)\cite{veintiuno}, the adquision system previously
used has been complemented by recording also the pulse shapes of
each event before and after amplification by mean of two 800 MHz
LeCroy 9362 digital scopes. These are analysed one by one by means
of a method based on wavelet techniques which allows us to access
the probability of this pulse to have been produced by a random
fluctuation of the baseline. This probability is used as a
criterium to reject events coming from electronic noise or
microphonics\cite{veintiocho}. According to the calibration of the
method, it works very efficiently for noise events above 4 keV.

\section{Results obtained from the IGEX set-ups \small{\cite{Mor00}\cite{veintiuno}}}

The detector IGEX-RG-II in both set-ups features an energy
threshold of 4 keV and an energy resolution of 0.8 keV at the
75~keV Pb x-ray line. We will show first the results obtained with
the set-up and shielding described above. They correspond to 30
days of analyzed data (Mt=60 kg-days) and they will be labelled
IGEX-2000 (Table \ref{tab-ig-10} and Fig.
\ref{dm-ig-2})\cite{Mor00}. The background rate recorded was $\sim
0.4$ c/(keV-kg-day) between 4--10~keV, $\sim 0.12$ c/(keV-kg-day)
between 10--20~keV, and $\sim 0.05$ c/(keV-kg-day) between
20--40~keV. Fig.~\ref{dm-ig-2} shows the RG-II 30-day spectrum in
this shielding. The numerical data are given in Table
\ref{tab-ig-10}. The corresponding $\sigma$(m) exclusion plot is
given in Fig. 3 following the procedure explained below. Earlier
results from the COSME detector (COSME-1)\cite{Jmor}\cite{Gar92}
are included in Fig. 3 (dot-dashed line). The recent results
obtained in its current set-up
(COSME-2)\cite{Ceb00}\cite{dieciseis}, corresponding to the
spectrum of 311 days depicted in Fig. \ref{fig_1} and listed in
Table \ref{tab-cos-9} are also shown in its corresponding
$\sigma$(m) exclusion contour of Fig. 3 (thick-dashed line).

\begin{figure}[ht]
\centerline{ \epsfxsize=10cm \epsffile{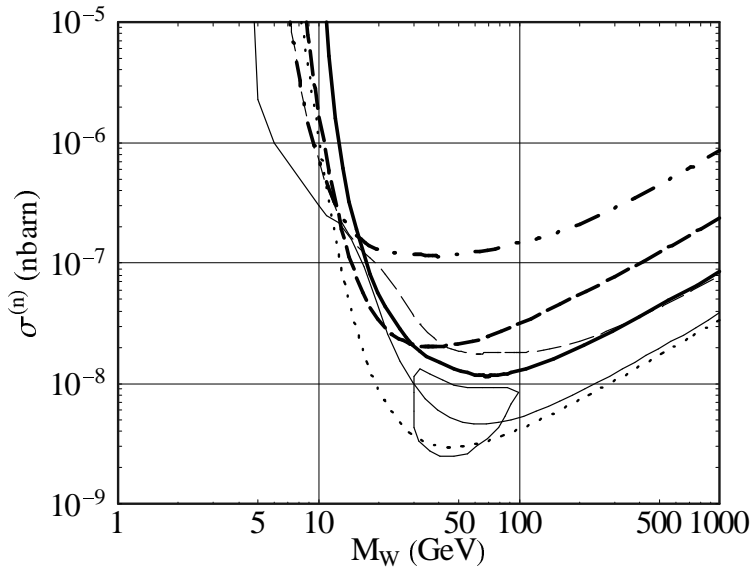} }
\caption{IGEX-DM exclusion plot for spin-independent interaction
obtainedfrom Canfranc IGEX-2000 Ge detector (thick solid line).
Results obtained in other Germanium experiments are also shown:
Canfranc COSME-1 data (dot-dashed line), recent COSME-2 data
(thick dashed line), and the previous Ge-combined bound (thin
dashed line) obtained for the other Ge experiments ---including
the last Heidelberg-Moscow data. The result of the DAMA NaI-0
experiment (thin solid line) is also shown. The ``triangle'' area
corresponds to the (3$\sigma$) annual modulation effect reported
by the DAMA collaboration (including NaI-1,2,3,4 runnings). The
IGEX-DM projection (dotted line) is shown for 1 kg-year of
exposure with a background rate of 0.1 c/(keV-kg-day).}

\label{dm-ig-4}
\end{figure}

\begin{table}[htb]
\begin{center}
\begin{tabular}[h]{cccccc}
\hline \multicolumn{1}{r}{{\bf E (keV)}}
                 & \multicolumn{1}{r}{{\bf counts}}

                 & \multicolumn{1}{r}{{\bf E (keV)}}
                 & \multicolumn{1}{r}{{\bf counts}}

                 & \multicolumn{1}{r}{{\bf E (keV)}}
                 & \multicolumn{1}{r}{{\bf counts}}
                  \\
\hline
\small
4.5 &   18  &   19.5    &   4   &   34.5    &   4   \\ 5.5 &   25
&   20.5    &   5   &   35.5    &   4   \\ 6.5 &   16  &   21.5 &
1   &   36.5    &   6   \\ 7.5 &   11  &   22.5    &   4   & 37.5
&   3   \\ 8.5 &   23  &   23.5    &   4   &   38.5    & 3   \\
9.5 &   9   &   24.5    &   4   &   39.5    &   3   \\ 10.5 &   12
&   25.5    &   4   &   40.5    &   5   \\ 11.5    &   17 &   26.5
&   4   &   41.5    &   4   \\ 12.5    &   12  & 27.5    &   9   &
42.5    &   0   \\ 13.5    &   7   &   28.5 &   4   &   43.5    &
2   \\ 14.5    &   6   &   29.5    &   3 &   44.5    &   3   \\
15.5    &   6   &   30.5    &   2   & 45.5    &   5   \\ 16.5    &
8   &   31.5    &   2   &   46.5 &   2   \\ 17.5    &   6   & 32.5
&   1   &   47.5    &   3
\\ 18.5    &   1   &   33.5    &   1   &   48.5    &   4   \\

\hline
\end{tabular}
\caption{Low-energy data from the IGEX RG-II detector (Mt $=$
80~kg~d) (IGEX-2001).} \label{tab-ig-11}
\end{center}
\end{table}

The exclusion plots resulting from the IGEX-2000 data are derived
from the recorded spectrum Fig. 2 in one-keV bins from 4~keV to
50~keV. The method followed in deriving the plot has been the same
for all the detectors. As recommended by the Particle Data Group,
the predicted signal in an energy bin is required to be less than
or equal to the (90\% C.L.) upper limit of the (Poisson) recorded
counts. The derivation of the interaction rate signal supposes
that the WIMPs form an isotropic, isothermal, non-rotating halo of
density $\rho = 0.3$~GeV/cm$^{3}$, have a Maxwellian velocity
distribution with $\rm v_{\rm rms}=270$~km/s (with an upper cut
corresponding to an escape velocity of 650~km/s), and have a
relative Earth-halo velocity of $\rm v_{\rm r}=230$~km/s. The
cross sections are normalized to the nucleon, assuming a dominant
scalar interaction. The Helm parameterization\cite{Eng91} is used
for the scalar nucleon form factor, and the recoil energy
dependent ionization yield used is the same that in Ref
\cite{Bau99} $\rm E_{vis} = 0.14 (E_{REC}) ^{1.19}$. The exclusion
plots obtained from the COSME-2 spectrum (Fig. \ref{fig_1} and
Table \ref{tab-cos-9}) and from COSME-1 (Fig. 1) have been
calculated, as previously noted, in the same way as for IGEX. As
shown in Fig. 3, the IGEX-2000 results\cite{Mor00} (thick-solid
line) exclude WIMP-nucleon cross-sections above 1.3x10$^{-8}$ nb
for masses corresponding to the 50 GeV DAMA region. Also shown is
the combined germanium contour (thin-dashed line), including the
last Heidelberg-Moscow data\cite{Bau99} (recalculated from the
original energy spectra with the same set of hypotheses and
parameters), the DAMA experiment contour plot derived from Pulse
Shape Discriminated spectra\cite{trece}, and the DAMA region
corresponding to their reported annual modulation
effect\cite{Ber99}. The IGEX-2000 exclusion contour\cite{Mor00}
improves that of other Germanium experiments for masses between 20
GeV and 200 GeV, which includes the mass region corresponding to
the neutralino tentatively assigned to the DAMA modulation effect
and results from using only raw data without background
subtraction. The COSME-2\cite{Ceb00}\cite{dieciseis} exclusion
contour also slightly improves the Ge-combined plot for masses
between 20 and 40 GeV.

\begin{figure}[ht]
\centerline{ \epsfxsize=10cm \epsffile{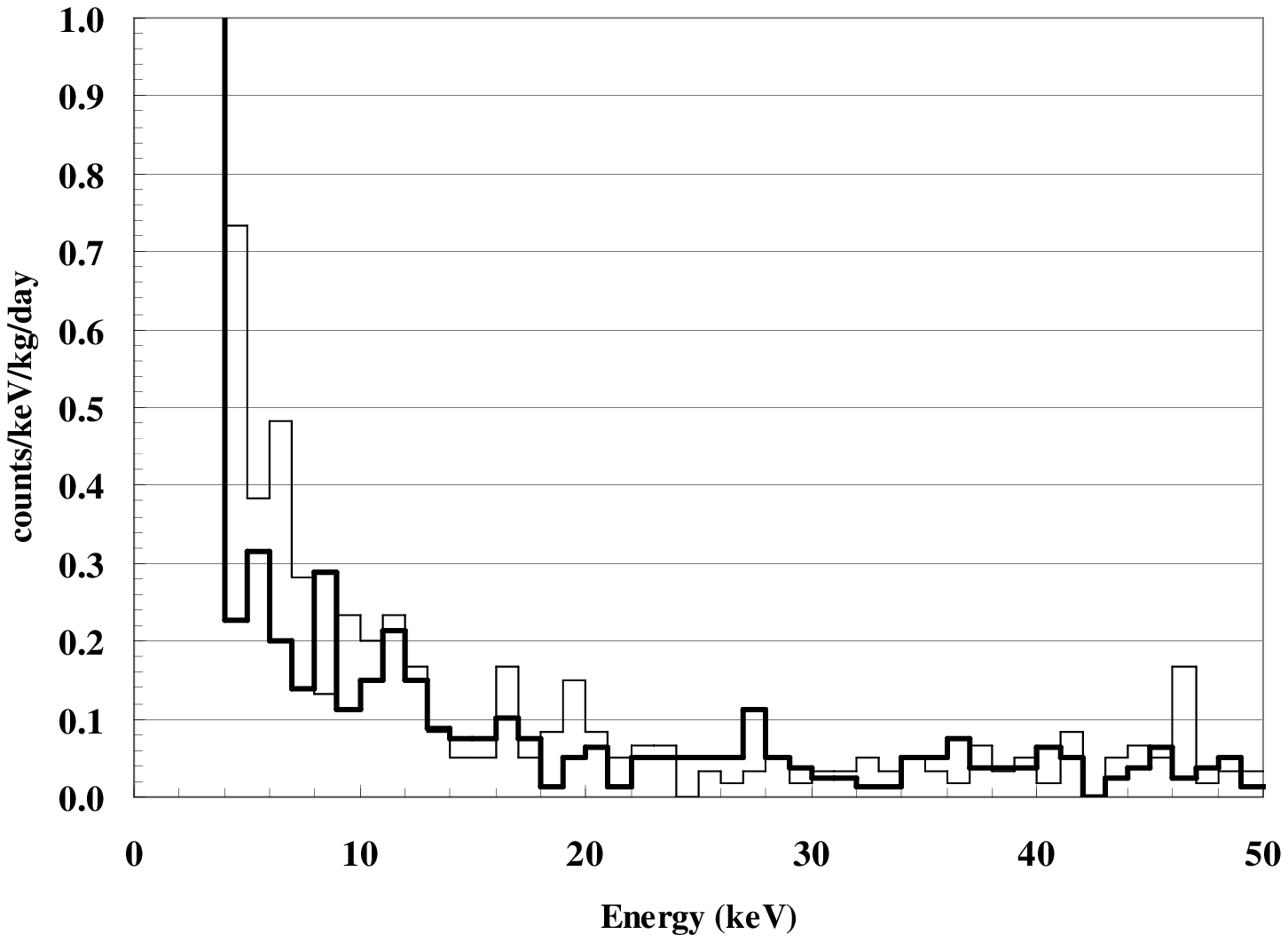} }
 \caption{Normalized low energy spectrum of the IGEX RG-II detector
 corresponding to the 80~kg~d (IGEX-2001) (thick line)
 compared to the previous 60~kg~d spectrum (IGEX-2000).}
 \label{dm-ig-04bis}
\end{figure}

Then, a new set-up was constructed, with the purpose of further
improving the background. In particular all the detectors except
IGEX RG-II were removed from the shielding, leaving RG-II alone. A
new, thorough cleaning of the inner layers of the bricks of lead
was performed, and the free space left was stretched. In this new
set-up, the detector RG-II is surrounded by not less than 40-45 cm
of lead of which 25 cm are archaeological. Also the muon veto
covers now more completely the ensemble, because three dewars have
been removed. A substantied improvement is that of the neutron
shielding which has been enlarged to up to 40 cm of thickners ,
consisting of polyethylene blocks and borated water tanks.

\begin{figure}[ht]
\centerline{ \epsfxsize=10cm \epsffile{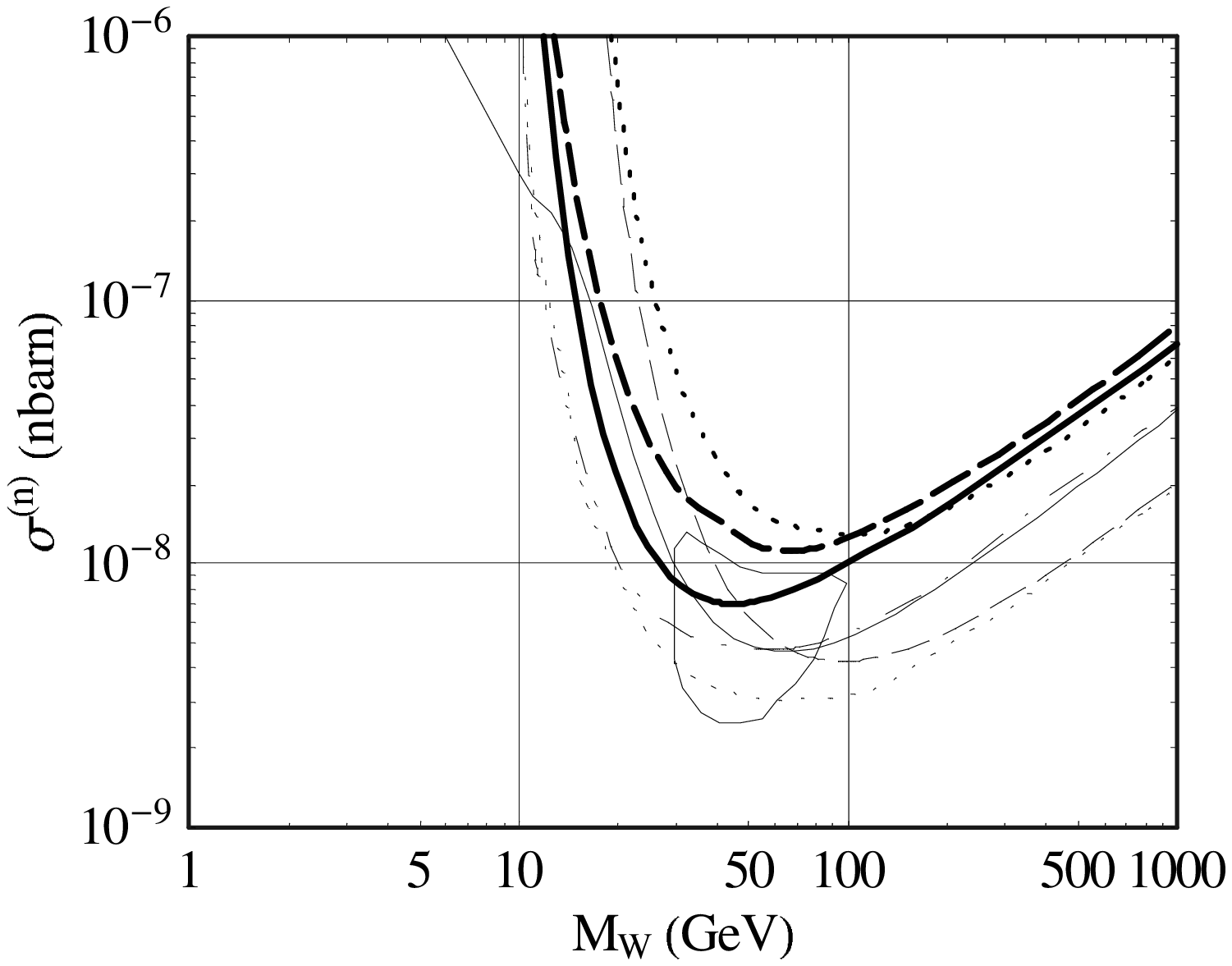} }
 \caption{IGEX-DM 2001 exclusion plot for spin-independent
interaction obtained in this work (thick solid line) compared with
the previous exclusion obtained by IGEX-DM 2000 (dashed thick
line) and the last result obtained by the Heidelberg-Moscow
germanium experiment cite{Bau} (dotted line) recalculated from the
original spectrum with the same set of hypothesis and parameters.
The closed line corresponds to the (3$\sigma$) annual modulation
effect reported by the DAMA collaboration (including NaI-1,2,3,4
runnings). The thin solid line is the exclusion line obtained by
DAMA NaI-0 by using Pulse Shape Discrimination. The two other
experiments which have entered the DAMA region are also shown:
EDELWEISS (thin dashed line) and the CDMS exclusion contour (thin
dotted line) and its expected sensitivity (thin dot-dashed line).
%Also
%shown are the other three experiments which enters in the DAMA
%region: CDMS \cite{Abusaidi:2000} (dot-dashed thin line),
%EDELWEISS \cite{Benoit:2001} (dashed thin line) and DAMA NaI-0
%\cite{Ber96} (thin solid line).
} \label{dm-ig-5}
\end{figure}

\begin{figure}[ht]
\centerline{ \epsfxsize=7.5cm \epsffile{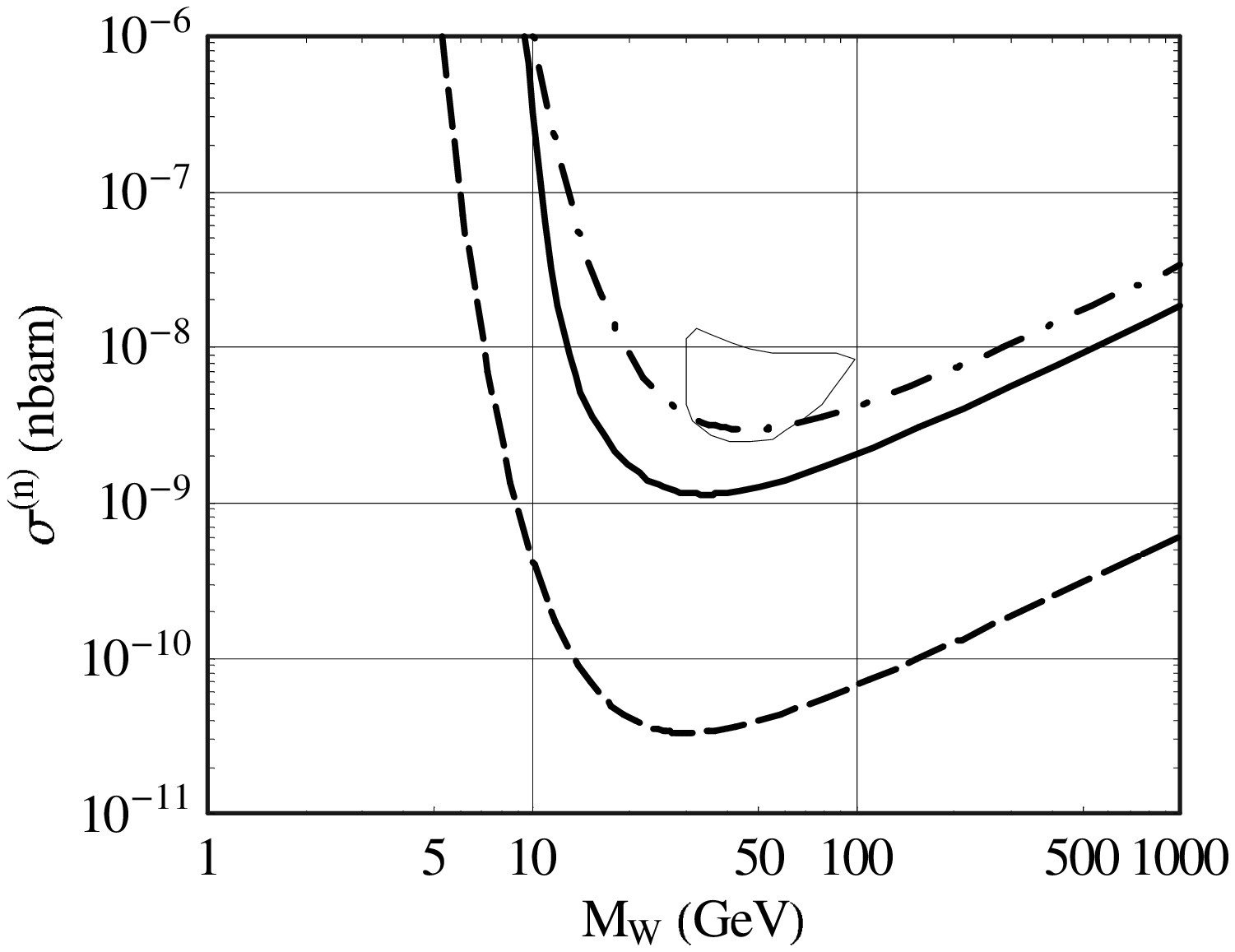} }
 \caption{IGEX-DM projections are shown for
 a flat background rate of 0.1~c/keV/kg/day (dot-dashed line) and 0.04~c/keV/kg/day (solid line) down
 to the threshold at 4 keV, for 1~kg~year of exposure.
 The exclusion contour expected for GEDEON is also
 shown (dashed line) as explained in the text.} \label{dm-ig-6}
\end{figure}

The results obtained whit this new shielding are from a recent run
and correspond to an exposure of Mt=80 kg~days\cite{veintiuno}.
The spectrum obtained in this new set-up, and labelled IGEX-2001,
is shown in Figure 4 compared with the previous IGEX-2000
spectrum\cite{Mor00}. The numerical data are given in Table 3.

\begin{figure}[h]
\centerline{ \epsfxsize=7.5cm \epsffile{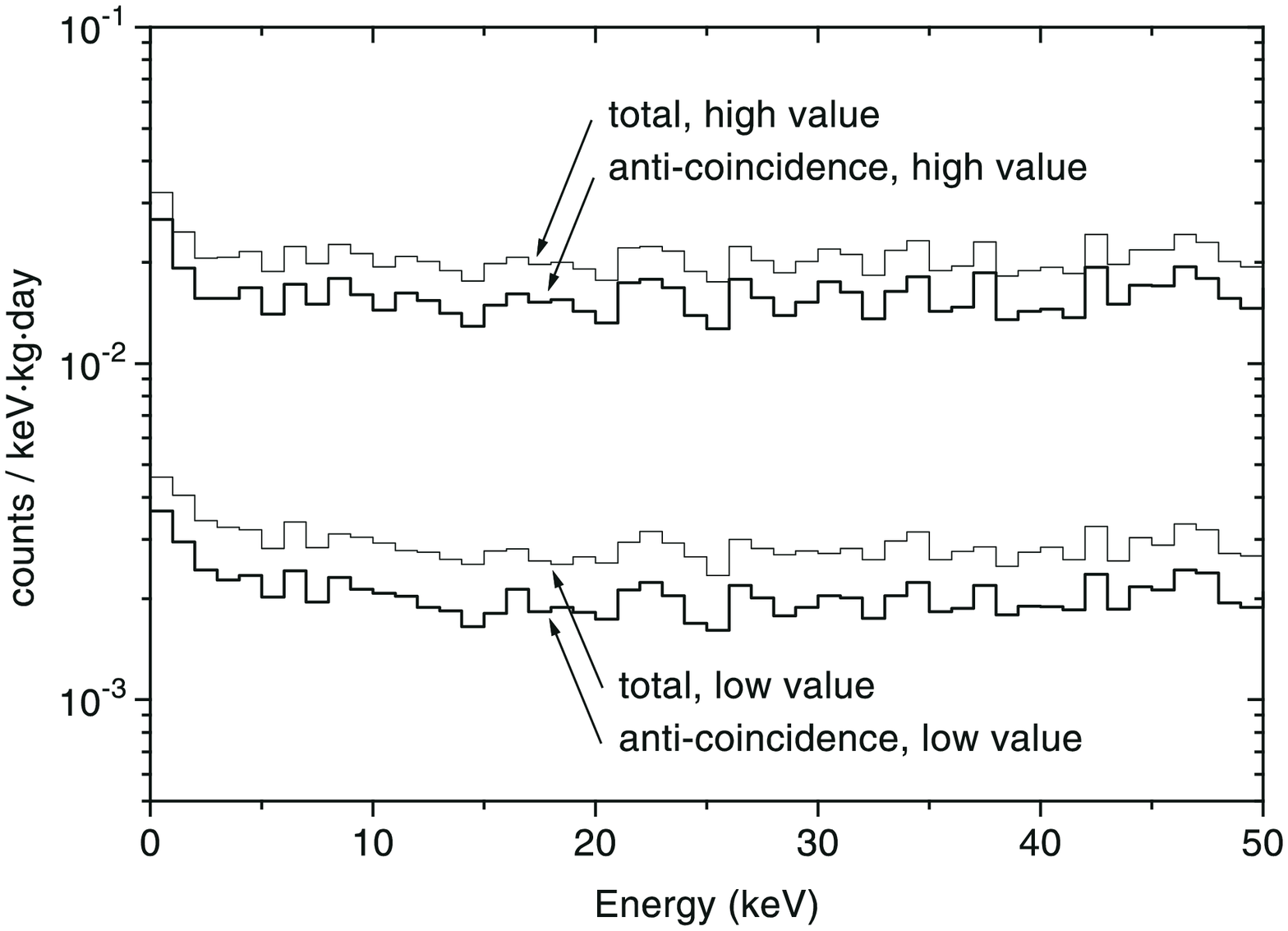} }
 \caption{Montecarlo estimaged background of the GEDEON detector
 project (energy interval 0---50 keV).}
  \label{fig7}
  \end{figure}

The energy threshold of the detector is, as in previous runnings,
4 keV and the FWHM energy resolution at the 75 keV Pb X-ray line
is 800 eV. The background rate recorded was $\sim 0.21$
c/keV/kg/day between 4--10~keV, $\sim 0.10$ c/keV/kg/day between
10--20~keV, and $\sim 0.04$ c/keV/kg/day between 25--40~keV. As it
can be seen, the background below 10 keV has been substantially
reduced (about a factor 50\%) with respect to that obtained in the
previous set-up of IGEX-2000, essentially due to the improved
shielding (both in lead and in polyethylene-water).This suggests
that the neutrons could be an important component of the low
energy background in IGEX\cite{veintiocho}.

The new exclusion plot derived in this improved conditions
IGEX-2001\cite{veintiuno} is shown in Fig.~\ref{dm-ig-5} (thick
solid line). It improves the IGEX-2000\cite{Mor00} exclussion
contourn (thick dashed line) as well as that of the other previous
germanium ionization experiments (and in particular that of the
last result of Heidelberg-Moscow experiment\cite{Bau99} now
specifically depicted by the thick dotted line) for a mass range
from 20~GeV zone to 200~GeV, which encompass, as already said,
that of the DAMA mass region. In particular, this new IGEX result
excludes WIMP-nucleon cross-sections above 7 $\times 10^{-6}$ pb
for masses of 40-60 GeV and enters the DAMA region. IGEX excludes
the upper left part of this region. That is the first time that a
direct search experiment without background discrimination, but
with very low (raw) background, enters such region. Also shown for
comparison are the contour lines of the other experiments,
CDMS\cite{Abusaidi:2000} and EDELWEISS\cite{Benoit:2001} (thin
dashed line), which have entered that region by using bolometers
which also measure ionization. The DAMA region (closed line)
corresponding to the $3\sigma$ annual modulation effect reported
by that experiment\cite{Ber99} and the exclusion plot obtained by
DAMA NaI-0\cite{trece} (thin solid line) by using statistical
pulse shape discrimination are also shown. A remark is in order:
for CDMS two contour lines have been depicted according to a
recent recommendation, the exclusion plot published in Ref.
\cite{Abusaidi:2000} (thin dotted line) and the CDMS expected
sensitivity contour (thin dot-dashed line)\cite{Sadoulet:2001}.

Data collection is currently in progress and some strategies are
being considered to further reduce the low energy background.
Another 50 \% reduction from 4 keV to 10 keV (which could be
reasonably expected) would allow to explore practically all the
DAMA region in 1~kg~y of exposure. In the case of reducing the
background down to the flat level of 0.04 c/kg/keV/day (currently
achieved by IGEX for energies beyond 20 keV), the DAMA region
would be widely surpassed. In Figure~\ref{dm-ig-6} we plot the
exclusions obtained with a flat background of 0.1 c/kg/keV/day
(dot-dashed line) and of 0.04 c/kg/keV/day (solid line) down to
the current 4 keV threshold, for an exposure of 1~kg~year. As can
be seen, the complete DAMA region (m=$52^{+10}_{-8}$ GeV,
$\sigma^p$=($7.2^{+0.4}_{-0.9}$)x10$^{-9}$ nb) could be tested
with a moderate improvement of the IGEX performances.

\centerline {\includegraphics[height=9cm]{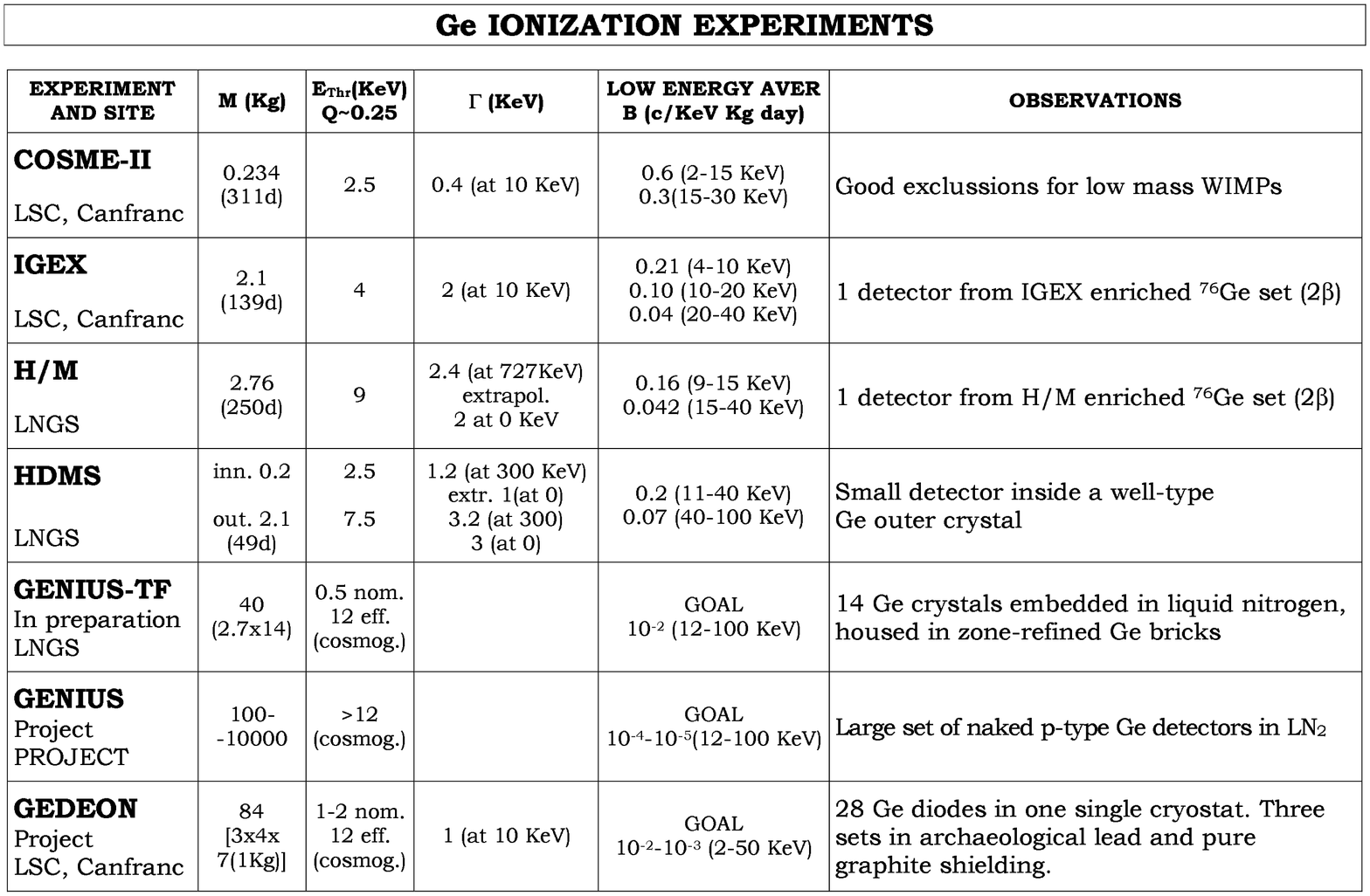}}
\centerline{Table5}

A new experimental project on WIMP detection using larger masses
of Germanium of natural isotopic abundance (GEDEON, GErmanium
DEtectors in ONe cryostat) is planned\cite{Mor992}. It will use
the technology developed for the IGEX experiment and it would
consist of a set of $\sim$1 kg germanium crystals, of a total mass
of about 28 kg, placed together in a compact structure inside one
only cryostat. This approach could benefit from anticoincidences
between crystals and a lower components/detector mass ratio to
further reduce the background with respect to IGEX.

The GEDEON single cell is a cylindrical cryostat in electroformed
copper (dimensions 20 cm diameter $\times$ 32 cm long) hosting 28
germanium crystals which share the same common copper cryostat
(0.5 mm thick). The Ge crystals, are arranged in four plates
suspended from copper rods. The cell is embedded into a
precision-machined hole made in a Roman lead block providing a
shield of 20 cm, and surrounded by another lead shielding 20 cm
thick. A cosmic veto and a large neutron shield complete the
shielding.

The preliminary MC estimated background in the $1 \sim 50$ keV
region ranges from $2 \times 10^{-2}$ to $2 \times 10^{-3}$ c/keV
kg day, according to the level of radioimpurities included as
input\cite{veintinueve}. The radiopurity assays have been carried
out in the Canfranc Laboratory for the lead and copper components
of the shielding. The background final goal of GEDEON, below 100
keV, would be hopefully in the region of $10^{-3}$ c/keV kg day
and this value has been used to calculate anticipated $\sigma$(m)
exclusion plots in the most favourable case. The expected
threshold assumed has been $\rm E_{\rm thr}=2$ keV and the energy
resolution in the low energy region has been taken $\Gamma \sim 1$
keV. The MC estimated background of the GEDEON unit cell (28
crystals) is given in Fig. 7. A detailed study is in progress to
assess the physics potential of this device. The exclusion plot
which could be expected with such proviso for 24 kg y of exposure
is shown in the Figure~\ref{dm-ig-6}. Moreover, following the
calculations presented in\cite{veintiseis}, GEDEON would be
massive enough to search for the WIMP annual modulation
effect\cite{Dru86} and explore positively an important part of the
WIMP parameter space including the DAMA region. A second phase of
GEDEON with four cryostats (112 detectors and a total mass of 92
kg of Ge) is also being considered.

\section*{Acknowledgments}

The data presented here regarding Ge ionization detectors resulted
from collaborative research with the members of the COSME and the
IGEX collaborations, formed by C.E. Aalseth, F.T. Avignone, R.L.
Brodzinski, S. Cebri\'{a}n, E. Garc\'{\i}a, W.K. Hensley, I.G. Irastorza,
I.V. Kirpichnikov, A.A. Klimenko, H.S. Miley, J. Morales, A. Ortiz
de Sol\'{o}rzano, S.B. Osetrov, V.S. Pogosov, J. Puimed\'{o}n, J.H.\
Reeves, M.L. Sarsa, A.A. Smolnikov, A.G. Tamanyan, A.A. Vasenko,
S.I. Vasiliev, J.A. Villar members to whom I am deeply indebted.
The results contained in this paper have already been published in
the open literature. I wish to thank especially S. Cebri\'{a}n and
I.G. Irastorza for their invaluable collaboration in the making of
the exclusion plots and to J. Morales and I.G. Irastorza for many
useful discussions on IGEX-DM. I thank S.Cebrian for allowing me
the use of the MC results for GEDEON and J.Puimedon for the making
of the radiopurity measurement. The present work was partially
supported by the CICYT and MCyT (Spain) under grant number
AEN99-1033.

\centerline{\includegraphics[height=18cm]{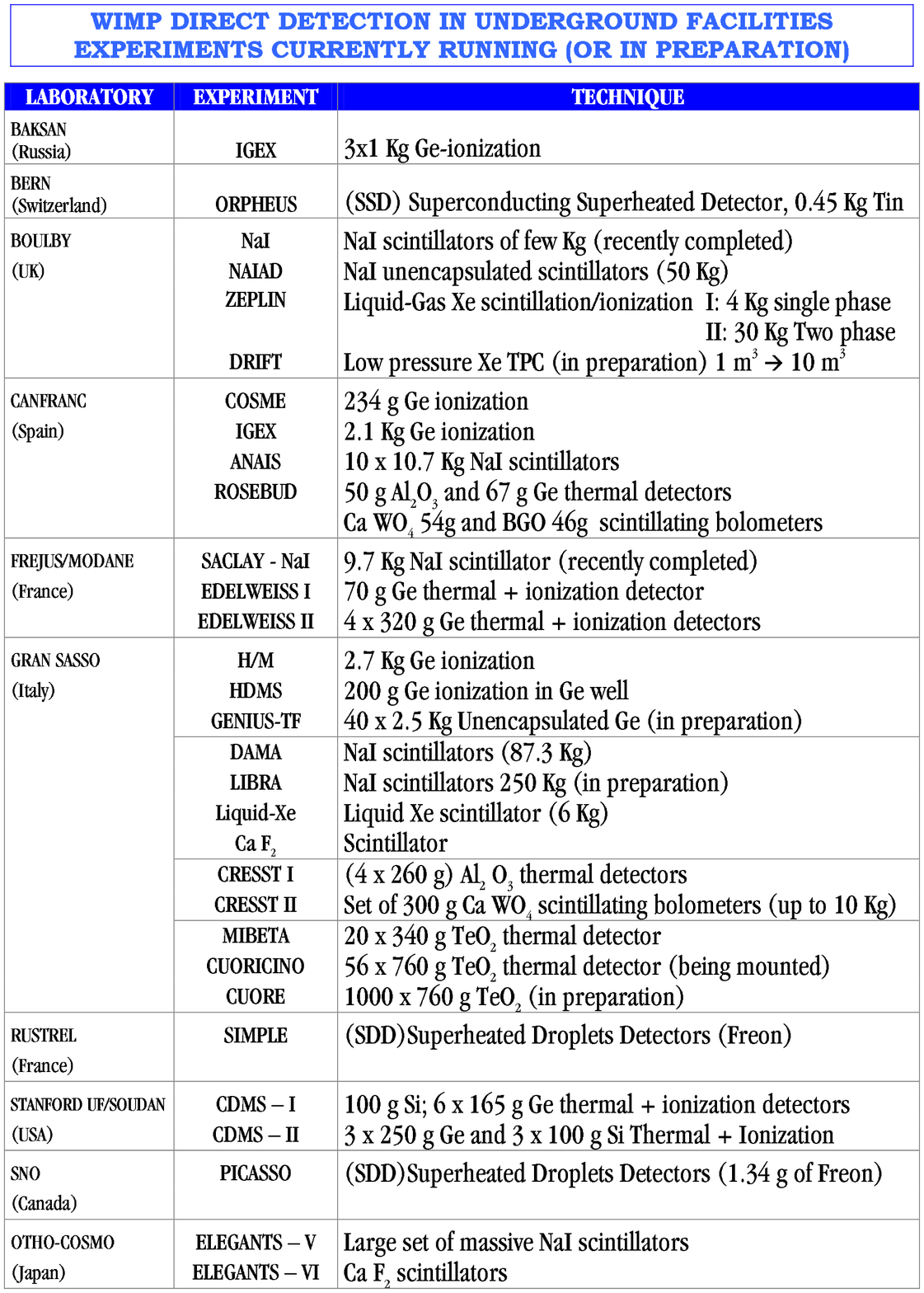}}
\centerline{Table4} 


\vspace{0.5cm}

\begin{thebibliography}{99}

\bibitem{Gri} G. Jungman, M. Kamionkowski and K. Griest, Phys.
Rep. {\bf 267} (1996) 195.

\bibitem{Mor} A. Morales, ``Dark Matter and its Detection'', Summary Talk
given at the NUPECC Workshop on Present and Future of Neutrino
Physics, Frascati, NUPECC Report in Highlights and Opportunities
in Nuclear Physics, Ed. by J. Vervier et al., December 1997. ({\tt
astro-ph/9810341}).

\bibitem{Mor2} F.T. Avignone and A.Morales, Proc. Int. Conference on
Neutrino Physics and Astrophysics. Helsinki, June 1996, ed. K.
Enkvist et al. in World Scientific Pub. (1997) p. 413; A. Morales
``Selected Projects in Direct Detection of Dark Matter'', Proc.
Neutrino Telescopes Workshop, Venice, February 1999. Ed. M.
Baldo-Ceolin, p. 24; L. Baudis and H.V. Klapdor "Direct Detection
of Non Baryonic Dark Matter", astro-ph /0003434 and Y. Ramachers
"Non Baryonic Dar Matter Searches"; XI Rencontres de Blois, June
1999, astro-ph /9911260.

\bibitem{Mor3} A. Morales, ``Direct Detection of WIMP Dark Matter''
({\tt astro-ph/9912554}). Review Talk at the TAUP 99 Workshop,
College de France, Paris. Nucl. Phys. B (Proc. Suppl.) 87 (2000)
477 and Review Talk at the TAUP 2001 Workshop, Laboratori
Nazionali del Gran Sasso, Italy, Sept. 2001, to be published in
Nucl. Phys. B (Proc. Suppl.) 2002.

\bibitem{taup} Topics in Astroparticles and Underground Physics,
TAUP 1999 Proceedings, Nucl. Phys B (Proc. Suppl.) 87 (2000),
Edited by J.Dumanchez, M.Froissart and D.Vignaud.

\bibitem{Ceb00} S. Cebri\'{a}n et al., New Journal of Physics {\bf 2} (2000)
({\tt http://www.njp.org}).

\bibitem{siete} S.P. Ahlen et al., Phys. Rev. D 33 (1987) 603. and A.K.
Drukier et al., Nucl. Phys. B (Proc. Suppl.) 28A (1992) 293.

\bibitem{Jmor} J. Morales et al., Nucl. Instrum. \& Meth. {\bf
A321} (1992) 410.

\bibitem{Gar92} E. Garc\'{\i}a et al., Nucl. Phys. B (Proc. Suppl.) {\bf
28A} (1992) 286 and Phys. Rev. {\bf D51} (1995) 1458.

\bibitem{diez} D.O. Caldwell et al., Phys. Rev. Lett. 61 (1988)
510.

\bibitem{once} D. Reusser et al., Phys. Lett. B 255 (1991) 143.

\bibitem{doce} M. Beck et al., Phys. Lett. B 336 (1994) 141.

\bibitem{trece} R. Bernabei et al., Phys. Lett. {\bf B379} (1996) 299.

\bibitem{Aal} C. Aalseth et al., Phys. Rev. {\bf C59} (1999)
2108.

\bibitem{Gon99} D. Gonz\'{a}lez et al., Proc. TAUP 99 Workshop, Coll\`{e}ge de
France, Paris, Nucl. Phys. B (Proc. Suppl.) {\bf 87} (2000) 278.

\bibitem{dieciseis} A. Morales. "WIMP searches at Canfranc with
Germanium Detectors". Proc. of South Carolina Symposium on
Neutrino Physics, Ed. H. Kuborera et al., World Scietific Pub.
2000.

\bibitem{Eng91} J. Engel, Phys. Lett. {\bf B264} (1991) 114.

\bibitem{Bau99} L. Baudis et al., Phys. Rev. {\bf D59} (1999) 022001.

\bibitem {Ber99} R. Bernabei et al., Phys. Lett. {\bf B450} (1999) 448
and Phys. Lett. B480 (2000) 23.

\bibitem{Mor00} A.~Morales et al. [IGEX Collaboration], Phys. Lett. B {\bf 489} (2000) 268 [hep-ex/0002053].

\bibitem{veintiuno}  A.~Morales et al. [IGEX Collaboration], hep-ex/0110061 oct.
2001 and I.G. Irastorza et al. (IGEX coll.), talk given at the
TAUP Workshop Sept. 2001, LN Gran Sasso. To be published in Nucl.
Phys. B (Proc. Suppl.) 2002.

\bibitem{Abusaidi:2000}
R.~Abusaidi et al. [CDMS Collaboration], Phys.\ Rev.\ Lett.\  {\bf
84} (2000) 5699 [astro-ph/0002471].

\bibitem{Benoit:2001}
A.~Benoit et al. [EDELWEISS Collaboration], Phys.\ Lett.\ B {\bf
513} (2001) 15 [astro-ph/0106094].

\bibitem{Sadoulet:2001}
B.~Sadoulet, private communication to A. Morales and talk given at
TAUP 2001 Workshop, LNGS (September 2001). To be published in
Nucl. Phys. B (Proc. Suppl.) 2002.

\bibitem{Mor992} A. Morales et al., GEDEON, a project for WIMP
searches with a set of natural abundance Ge diodes in a single
cryostat. Preliminary Study for Submission to CICYT (Spain),
January 1999.

\bibitem{veintiseis} S. Cebrian et al., Astropart. Phys. 14
(2001) 339.

\bibitem{Dru86} A.K. Drukier et al., Phys. Rev. {\bf D33} (1986) 3495.

\bibitem{veintiocho} See Igor G.Irastorza. Thesis Disertation.
Univ. of Zaragoza. April 2001.

\bibitem{veintinueve} S. Cebrian, private communication.

\bibitem{treinta} See for instance A.Bottino et al. DFTT 35/2000
and references therein; J. Ellis et al. hep-ph/0007113 and P.
Gondolo hep-ph/0008022.

\end{thebibliography}
\end{document}